\documentclass[aps,pra,showpacs,twocolumn]{revtex4}
\usepackage{graphicx}
\usepackage{epsfig}
\usepackage{dcolumn}
\usepackage{bm}

\begin{document}

\title{Density wave instability in a 2D dipolar Fermi gas}

\date{\today}

\author{Yasuhiro Yamaguchi}
\affiliation{Research Center for Nuclear Physics, Osaka University, Ibaraki, Osaka 567-0047, Japan}
\author{Takaaki Sogo}
\affiliation{Institut f$\ddot{u}$r Physik, Universit$\ddot{a}$t Rostock, D-18051 Rostock, Germany}
\author{Toru Ito}
\affiliation{Department of Physics, Faculty of Science, Tokyo University of Science,
1-3 Kagurazaka, Shinjuku, Tokyo, 162-8601, Japan}
\author{Takahiko Miyakawa}
\affiliation{Faculty of Education, Aichi University of Education, Hirosawa 1, Igaya-cho, Kariya 448-8542, Japan}
\email{takamiya@auecc.aichi-edu.ac.jp}

\begin{abstract}
We consider a uniform dipolar Fermi gas in two-dimensions (2D) where the dipole moments of fermions
are aligned by an orientable external field.
We obtain the ground state of the gas in Hartree-Fock approximation and
investigate RPA stability against density fluctuations of finite momentum.
It is shown that the density wave instability takes place in a broad region where the system
is stable against collapse.
We also find that the critical temperature can be a significant fraction of Fermi temperature
for a realistic system of polar molecules.
\end{abstract}


\maketitle

\section{Introduction}

The realization of ultracold molecules confers remarkable opportunities to study new states of quantum matter, and is also of much interest for quantum computing~\cite{Carr09}.
In particular, the investigation of polar molecules is a good candidate for studying various quantum many-body states because the anisotropic and long-range nature of the dipole-dipole interaction offers rich properties that do not occur in non-dipolar systems~\cite{BaranovReview}. 

Recently K.-K.~Ni {\it et al.} succeeded in the creation of a dense gas of $^{40}$K-$^{87}$Rb polar molecules
with temperature $T\approx 2.5T_F$~\cite{Ni08}, where $T_F$ is the Fermi temperature.
For the absolute rovibrational ground state of the fermionic molecule
attained in the experiment, the magnitude of the electric dipole moment is $0.566$ Debye,
yielding to significant interaction effects in a degenerate gas of dipolar fermions at lower temperatures.
As shown in Ref.~\cite{Miyakawa08}, the anisotropy of the dipole-dipole interaction results in a deformed Fermi surface
of the ground state of a dipolar Fermi gas.
Since the Fermi surface is the determining factor in low energy properties of the Fermi system,
theoretical studies of collective oscillations~\cite{Sogo09,Tohyama09}, stability~\cite{Zhang09}, 
expansion dynamics after turning off the trapping potential~\cite{Sogo09,Nishimura09},
zero sound propagation in a homogeneous system~\cite{Ronen09,Chan09},
and equilibrium properties at finite temperatures~\cite{Zhang10,Kestner10,Endo10}
under the deformed Fermi surface have been done recently.

Some of most fascinating challenges of a dipolar Fermi gas concern phase transitions of possible ordered phases.
So far, the realization of superfluid phase in 3D~\cite{BaranovSF,Zhao09} and 2D~\cite{Bruun08,Cooper09},
biaxial nematic phase~\cite{FregosoNJP}, and ferromagnetic phase of two component mixtures~\cite{FregosoPRL} 
have been proposed theoretically.
Moreover, there will be another interesting ordered phase arisen from the repulsive part of the long-range dipole-dipole force,
that is a density wave phase.

In a density wave phase, translational invariance is broken spontaneously, resulting in a distorted density distribution~\cite{Frohlich54}.
The phase transition of the charge density wave in 1D conductor was originally discussed by Peierls~\cite{Peierls},
where he found the instability of a metallic state at zero temperature.
In mid 1970s, the charge and spin density waves were discovered in experiments and
a number of interesting phenomena were found in static and dynamic properties of the density waves~\cite{Gruner}.

In the present paper, we consider a uniform dipolar Fermi gas in 2D with the dipole moments
aligned to an orientable external field.
We study a realization of the phase transition of a density wave
by evaluating the stability condition for the thermal equilibrium state in random phase approximation (RPA).
By calculating the critical temperature $T_c$, we will show that the density wave can be achieved at
a significant fraction of $T_F$ for a realistic system of polar molecules.
Here we note that the density wave is brought purely by the interaction effect
in contrast to the density waves in an optical lattice potential as discussed in Refs.~\cite{Quintanilla09,Lin09}.

This paper is organized as follows.
Section II presents the mean field model of the uniform gas of fermions interacting via dipole-dipole forces in 2D.
Section III presents the ground state properties at zero temperature and
discusses RPA stability.
We obtain a phase diagram at zero temperature as a variance of the magnitude of the interaction strength
and the angle between the direction of the dipole moments and normal direction of the 2D-plane.
Section IV applies our model into dipolar Fermi gases at finite temperatures,
which leads to the critical temperature of the density wave phase transition.
Section V is a summary.

\section{Dipolar Fermi gas in 2D}

\begin{figure}
    \begin{center}
    \includegraphics[width=7cm]{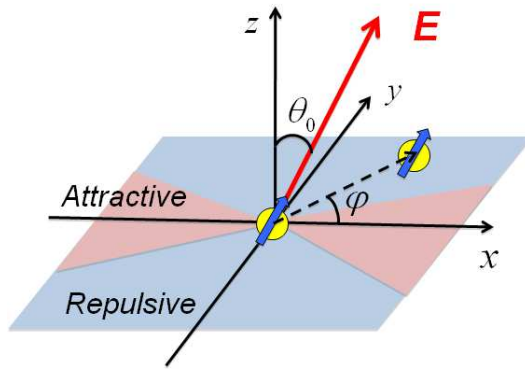}
    \end{center}
    \caption{(Color online) Experimental setup of 2D dipolar fermions. The dipole moments
    are alined by an orientable external field, ${\bm E}$, that forms the angle $\theta_0$ with respect to the $z$-axis.
    $\varphi$ denotes the azimuthal angle with respective to the $x$-axis of the relative coordinate
    of two particles in the $x$-$y$ plane.}
    \label{fig1}
\end{figure}

We consider a gas of dipolar fermions of mass $m$ and electric or magnetic dipole moment ${\bm d}$.
The dipoles are confined by a harmonic trapping potential $V(z)=m\omega_z^2 z^2/2$ 
with a trap frequency $\omega_z$ in the $z$-direction.
For $\hbar\omega_z \gg \epsilon_F$ where $\epsilon_F$ is the Fermi energy, the system is effectively 2D. 
The dipole moments are alined by an external electric or magnetic field, $\bm E$, 
subtending an angle $\theta_0$ with respect to the $z$-axis as shown in Fig.~\ref{fig1}.
In this case, the Hamiltonian of the system is given by
\begin{eqnarray}
\hat H=\sum_{i=1}^N -\frac{\hbar^2}{2m} \nabla_i^2
+\frac{1}{2}\sum_{i\neq j}V_{dd}(\bm{r}_i -\bm{r}_j),
\label{H}
\end{eqnarray}
where $N$ is the number of fermions and ${\bm r}_i$ is the position vector of $i$th-particle in the $x-y$ plane.
The 2D dipole-dipole interaction in Eq.~(\ref{H}) is described by
\begin{eqnarray}
\label{ddints}
V_{dd}(\bm{r})&=& \frac{d^2}{r^3}
\left\{1- 3\sin^2\theta_0\cos^2\varphi \right\}\, \nonumber \\
&=&\frac{d^2}{r^3}
\left\{P_2(\cos\theta_0)- \frac{3}{2}\sin^2\theta_0\cos2\varphi \right\},
\end{eqnarray}
where $\varphi$ is the azimuthal angle relative to the $x$-axis, see Fig.~\ref{fig1}.
In the above expression we have 2nd-order Legendre polynomial,
$P_2(\cos\theta_0)=(3\cos^2\theta_0-1)/2$.
According to the procedure introduced in Ref.~\cite{Chan09},
the momentum representation of the 2D dipole-dipole interaction of Eq.~(\ref{ddints})
is given by $V_{dd}({\bm q})=V_{0}+V_{2d}(\bm{q})$ where
\begin{eqnarray}
\label{intzero}
V_{0}&=&2\pi d^2P_2(\cos\theta_0)\frac{1}{r_c}, \\
\label{2dint}
V_{2d}(\bm{q})&=&\pi d^2q\left\{ -2P_2(\cos\theta_0)+ \sin^2\theta_0\cos2\phi \right\}.
\end{eqnarray}
In Eq.~(\ref{2dint}), $\phi$ is the angle with respect to the $q_x$-axis.
In Eq.~(\ref{intzero}), $r_c$ is a cut-off length and an order of size of a dipolar particle,
say size of a polar molecule, where the dipole moment is no longer ideal dipole moment.
Thus the divergence in the limit of $r_c \to 0$ is artificial~\cite{2dinteraction}.
As we shall see below, however, the total energy for dipolar fermions is always finite
even in the limit of $r_c \to 0$. Then we do not need to pay attention to such an artificial divergence.

The thermal equilibrium state of dipolar fermions at a temperature $T$
is assumed to be translational invariant in the $x$-$y$ plane.
In Hartree-Fock (HF) approximation, 
the total energy per area derived from the Hamiltonian of Eq.~(\ref{H}) 
is represented by
\begin{eqnarray}
\label{Etot}
E_{tot}&=&E_{kin}+E_{int}, \\
E_{kin}&=&\int \frac{d^2k}{(2\pi)^2}  \, \frac{\hbar^2 {\bm k}^2}{2m} f({\bm k}),
\label{ekin}\\
E_{int}&=&-\frac{1}{2} \int\frac{d^2k}{(2\pi)^2}  \int \frac{d^2k^\prime}{(2\pi)^2}\, 
f(\bm{k}) f(\bm{k}') V_{2d}(\bm{k} -\bm{k}'),  \label{eint}
\end{eqnarray}
where $f(\bm k)$ is the Fermi distribution function at $T$
\begin{equation}
\label{Fermidistribution}
   f(\bm k ) = \frac{1}{\exp\{(\epsilon(\bm k)-\mu)/k_BT\}+1},
\end{equation}
with Boltzman constant $k_B$ and the chemical potential $\mu$.
The single particle energy $\epsilon(\bm k)$ is obtained by solving the selfconsistent HF equation
\begin{equation}
\label{HFequation}
   \epsilon({\bm{k}}) =\frac{\hbar^2 \bm{k}^2}{2m}
   - \int \frac{d^2k^\prime}{(2\pi)^2} V_{2d}(\bm{k}-\bm{k}^\prime) f({\bm k}^\prime),
\end{equation}
under the constraint for the number density of fermions in 2D
\begin{equation}
\label{constraint}
   n_{2d}=\int \frac{d^2 k}{(2\pi)^2} f(\bm k).
\end{equation}

We note that the mean field interaction energy $E_{int}$ of Eq.~(\ref{eint}) is the sum of the direct
and exchange energy, where a total cancellation of the short range interaction $V_0$ of Eq.~(\ref{intzero})
occurs due to Fermi statistics.
Thus the divergence of $V_0$ in the limit of $r_c \to 0$ disappears in $E_{int}$ or $E_{tot}$.

\section{Ground state and stability at zero temperature}

In this section, we obtain the ground state at zero temperature $T=0$ in HF approximation
where the distribution function is replaced by
\begin{equation}
   f({\bm k}) = \Theta\left( \epsilon_F - \epsilon({\bm k}) \right),
\end{equation}
with $\mu=\epsilon_F$. In the above equation $\Theta()$ denotes Heaviside's step function.
When the external field tilts with respect to the $z$-axis, the anisotropy of $V_{2d}(\bm q)$
in the 2D momentum space causes the deformed Fermi surface.
Before the study of numerical calculations in HF approximation,
we introduce a variational approach that captures physical insight of the ground state.

\subsection{variational method}
In Refs.~\cite{Miyakawa08,Sogo09}, we developed a variational method (VM) to a 3D dipolar Fermi gas,
which describes the deformed Fermi surface with an ellipsoidal shape.
In Ref.~\cite{Bruun08}, Bruun {\it et al.} applied VM to a 2D dipolar Fermi gas with the variational density distribution
\begin{equation}
\label{defdis}
   f({\bm{k}})=\Theta\left(k_{F0}^2 - \alpha^2 k_x^2 -\frac{1}{\alpha^2}k_y^2\right),
\end{equation}
where the positive parameter $\alpha$ describes an elliptical Fermi surface
and $k_{F0}\equiv\sqrt{4\pi n_{2d}}$.
When $\theta_0 \neq 0$, the minus sign of the interaction energy~(\ref{eint}) and the anisotropy of 
$V_{2d}(\bm q)$ tend to stretch the Fermi surface along the $x$-axis leading to $\alpha<1$.

Under the assumption of the distribution function of Eq.~(\ref{defdis}),
the total energy in VM in units of $\hbar^2 n_{2d}^2/m$ is given by,
see Appendix~A, 
\begin{equation}
\label{varE}
   \mathcal{E}_{tot} = \frac{\pi}{2}\left( \frac{1}{\alpha^2}+ \alpha^2 \right)
   -\frac{32}{15} g I(\alpha; \theta_0),
\end{equation}
where $g\equiv 4md^2k_{F0}/3\pi\hbar^2$ and $I(\alpha;\theta_0)$ is defined by
\begin{eqnarray}
   I(\alpha;\theta_0) \equiv -\frac{2P_2(\cos\theta_0)}{\alpha}E(1-\alpha^4)\hspace{2.6cm} \nonumber\\
   +\frac{\sin^2\theta_0}{\alpha}\left\{-\frac{2\alpha^4}{1-\alpha^4}K(1-\alpha^4)
   +\frac{1+\alpha^4}{1-\alpha^4}E(1-\alpha^4)
   \right\}
   \label{deffunc}
\end{eqnarray}
for $\alpha<1$ and $I(\alpha;\theta_0)=-\pi$ for $\alpha=1$. 
In Eq.~(\ref{deffunc}), $K(m)$ and $E(m)$ are the complete elliptic integrals of the first and second kind defined by
\begin{eqnarray}
   K(m) &=& \int^{\pi/2}_0 d\theta\, \frac{1}{\sqrt{1-m\sin^2\theta}}, \\
   E(m) &=& \int^{\pi/2}_0 d\theta\, {\sqrt{1-m\sin^2\theta}},
\end{eqnarray}
respectively.
We find the VM ground state by minimizing the total energy (\ref{varE}) giving rise to
an optimized variational parameter $\alpha=\alpha_0$.

\subsection{Ground state properties}

\begin{figure}
    \begin{center}
    \includegraphics[width=7cm]{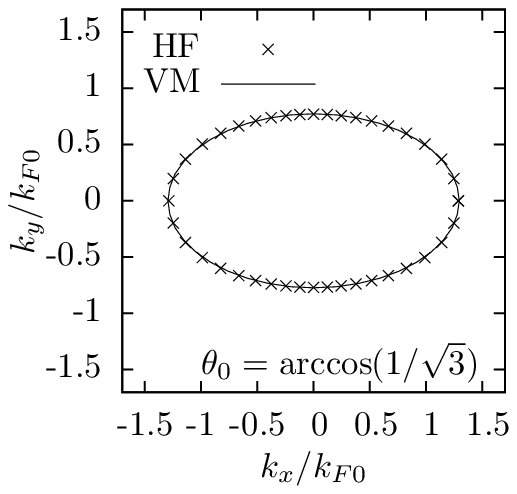}
    \end{center}
    \caption{Deformed Fermi surface for $g=1.0$ and $\theta_0=\arccos(1/\sqrt{3})$
    derived from numerical calculations in  HF approximation (cross) and VM (solid line).}
    \label{fig2}
\end{figure}

\begin{figure}
    \begin{center}
    \includegraphics[width=7cm]{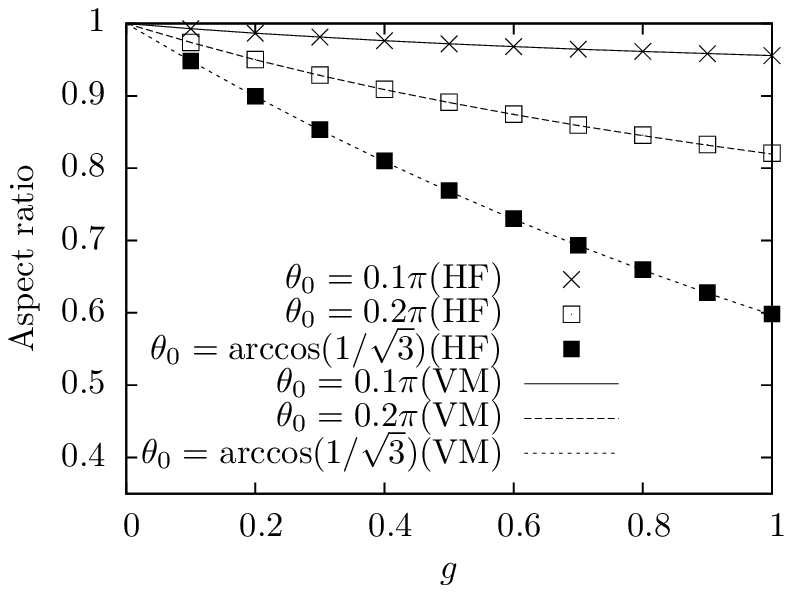}
    \end{center}
    \caption{Aspect ratio of the Fermi surface, $k_F(\pi/2)/k_F(0)$, as a function of $g$ for $\theta_0=0.1\pi$, 
    $\theta_0=0.2\pi$, and $\theta_0=\arccos(1/\sqrt{3})$ from top to bottom.
    The solid line, dashed line, and dotted line are results of the HF ground state,
    and the cross,  blank square, and filled square are results of the VM ground state.}
    \label{fig3}
\end{figure}

Figure~\ref{fig2} shows the deformed Fermi surface in the 2D momentum space in units of $k_{F0}$
for $g=1.0$ and $\theta_0=\arccos(1/\sqrt{3})$ where $P_2(\cos{\theta_0})=0$.
The cross and solid line correspond to results from numerical calculations in HF approximation and VM, respectively.
It is shown that two results match very well as the case for a 3D dipolar Fermi gas~\cite{Ronen09}
and the Fermi wave number derived from $\epsilon_F=\epsilon(\bm k)$, $k_F(\phi)$, is dependent on $\phi$.
In Fig.~\ref{fig3}, we plot the aspect ratio of the Fermi surface , $k_F(\pi/2)/k_F(0)$, 
of the HF (cross) and VM (solid line) ground state as a function of $g$ for $\theta_0=0.1\pi, 0.2\pi, \arccos(1/\sqrt{3})$.
We note that the aspect ratio in VM is given by $k_F(\pi/2)/k_F(0)=\alpha^2_0$, see Eq.~(\ref{defdis}).
This result reveals that the shape of Fermi surface is well approximated by the elliptical shape
for different values of $g$ and $\theta_0$.

\begin{figure}
    \begin{center}
    \includegraphics[width=7cm]{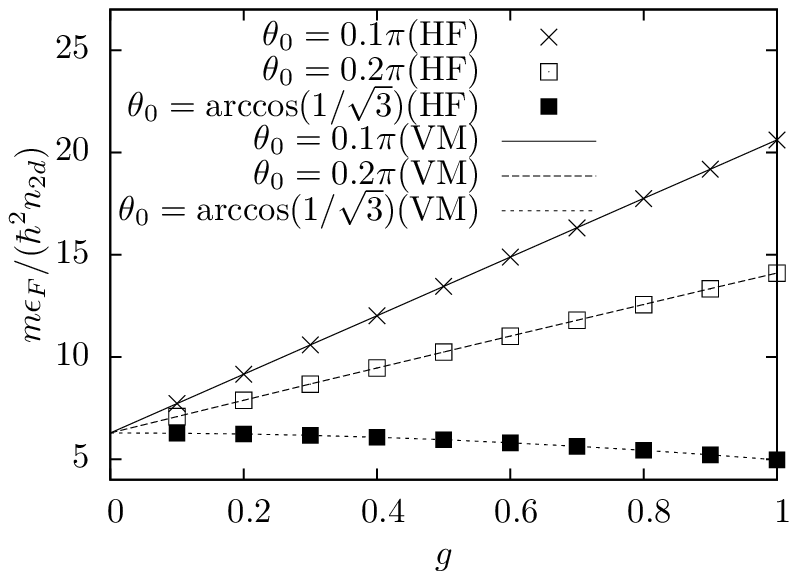}
    \end{center}
    \caption{Fermi energy as a function of $g$ for $\theta_0=0.1\pi$, 
    $\theta_0=0.2\pi$, and $\theta_0=\arccos(1/\sqrt{3})$ from top to bottom.
    The solid line, dashed line, and dotted line are results of the HF ground state,
    and the cross,  blank square, and filled square are results of the VM ground state.}
    \label{fig4}
\end{figure}

We numerically calculate the Fermi energy, $\epsilon_F$, by a partial derivative of the total energy of the system
with respect to the number density $n_{2d}$.
Figure~\ref{fig4} shows a plot of $\epsilon_F$ in units of $\hbar^2n_{2d}/m$
of the  HF (cross) and VM (solid line) ground state as a function of $g$ for different values of $\theta_0$.
Again two of results agree very well.
As $g$ increases, $\epsilon_F$ increases for $\theta_0=0.1\pi, 0.2\pi$ and decreases for $\theta_0=\arccos(1/\sqrt{3})$.
This result reveals that an average effect of $V_{2d}(\bm q)$
turns to be from repulsive to attractive as the direction of the external field is further apart from the $z$-axis.
%
%

\subsection{Stability condition}
We examine the stability of the ground state and obtain a phase diagram at zero temperature in the $g$-$\theta_0$ plane.
Owing to judge the stability of the system, we take RPA~\cite{Nozieres,noteRPA}
and obtain the stability condition
\begin{equation}
\label{RPAstability}
   1+V_{2d}({\bm q})\chi({\bm q}) > 0,
\end{equation}
where $\chi({\bm q})$ is the density-density response function against density fluctuations of
finite momentum $\bm q$ defined by
\begin{equation}
\label{response}
   \chi({\bm q}) = \int \frac{d^2 k}{(2\pi)^2}\, \frac{f({\bm k + \bm q})-f(\bm k)}
   {\epsilon({\bm k}) - \epsilon({\bm k+ \bm q})}.
\end{equation}

\begin{figure}
    \begin{center}
    \includegraphics[width=7cm]{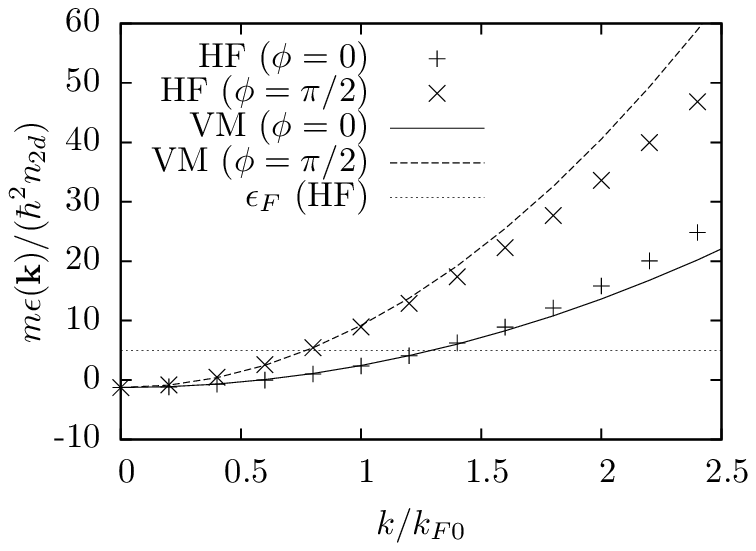}
    \end{center}
    \caption{Single particle energies $\epsilon(k,\phi=0)$ and $\epsilon(k,\phi=\pi/2)$
    for $g=1.0$ and for $\theta_0=\arccos(1/\sqrt{3})$
    derived from numerical calculations in HF approximation (crosses) and VM (solid and dashed lines).
    Dotted line represents the Fermi energy in HF approximation.}
    \label{fig5}
\end{figure}

As we have seen in the previous subsection, the aspect ratio and the Fermi energy in VM
approximate to those in HF approximation quite well.
Thus we calculate the response function by use of the single particle energy that reproduces
the variational distribution function~(\ref{defdis}). The single particle energy in VM reads
\begin{eqnarray}
\label{VMSPE}
   \epsilon(\bm k) &=& \epsilon(\bm 0) + \frac{\hbar^2}{2m}\lambda^2\left( \alpha_0^2k_x^2 +\frac{1}{\alpha_0^2}k_y^2 \right), \\
   \epsilon(\bm 0) &=& -\frac{\hbar^2 k_{F0}^2}{4m}g 
   I(\alpha_0,\theta_0),
\end{eqnarray}
where $\lambda$ represents the curvature of the single particle energy and
is determined by the relationship
\begin{equation}
   \epsilon_F = \epsilon(\bm 0) + \lambda^2 \frac{\hbar^2 k_{F0}^2}{2m}.
\end{equation}
Figure~\ref{fig5} shows the single particle energy in HF approximation and VM 
as a function of $k=|{\bm k}|$ at $\phi=0$ and $\pi/2$.
The single particle energy in VM matches with that in HF approximation below and near the Fermi energy.

\begin{figure}
    \begin{center}
    \includegraphics[width=7cm]{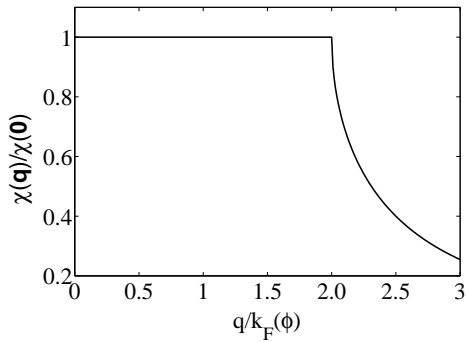}
    \end{center}
    \caption{Density-density response function in VM.}
    \label{fig6}
\end{figure}

By use of the density distribution~(\ref{defdis}) and single particle energy~(\ref{VMSPE}) in Eq.~(\ref{response}),
the analytic form of the response function for $\bm q=(q,\phi)$ is obtained by
\begin{equation}
\label{resfunction}
  \chi({\bm q})=
   \frac{m}{2\pi\lambda^2\hbar^2}\left[1-\sqrt{1-\left(\frac{2k_{F}(\phi)}{q}\right)^2}
   \Theta\left(q-2k_F(\phi)\right)\right]
\end{equation}
where the angle dependent Fermi wave number is given by
\begin{equation}
   k_{F}(\phi)=\frac{k_{F0}}{\sqrt{\alpha_0^2 \cos^2{\phi} + \frac{1}{\alpha_0^2}\sin^2{\phi} }}.
\end{equation}
We plot $\chi(\bm q)$ of Eq.(\ref{resfunction}) in Fig.~\ref{fig6} that is
constant below $q \leq2k_F(\phi)$ and monotonically decreasing for $q> 2k_F(\phi)$,
showing a singular behavior at $q=2k_F(\phi)$

Apart from $\theta_0=0$, the 2D dipolar interaction $V_{2d}(\bm q)$ is anisotropic and is most negative at
$\phi=\pi/2$ for a fixed magnitude of momentum $q$. Thus the system
is expected to be unstable against density fluctuations of the momentum ${\bm q} =(q,\pi/2)$.
Once the condition
\begin{equation}
\label{condition}
   V_{2d}\left(q,\frac{\pi}{2}\right) \chi\left(q,\frac{\pi}{2}\right) \leq -1
\end{equation}
is fulfilled, the system becomes unstable.
The minimum value of the left hand side of Eq.~(\ref{condition}) occurs at $q=2k_F(\pi/2)=2\alpha_0k_{F0}$
Thus we obtain the instability condition of the system as
\begin{equation}
\label{instcond}
   g \cos^2\theta_0\frac{\alpha_0}{\lambda^2} \geq \frac{2}{3\pi} = 0.212.
\end{equation}
In case the above condition is satisfied, the density fluctuation of the momentum $q=2\alpha_0k_{F}$
along the $y$-axis starts to develop. This indicates that the phase transition from a normal phase gas into
a density wave phase takes place for larger $g$ and smaller $\theta_0$.
Note that the expected density wave is a planar wave in the transverse direction to the dipolar direction.
As shown in the instability condition (\ref{instcond}), the system becomes unstable
even when $\theta_0=0$ corresponding to the circular symmetric system.
In this case the density wave of a spherical wave is possible to appear as a stable ground state.

In occurrence of the density wave instability there are two key ingredients, 
that is, the linear momentum dependence of $V_{2d}(\bm q)$ and 
the singular behavior  of $\chi\left(q,\pi/2\right)$ at $q=2\alpha_0 k_{F0}$.
Thus the present mechanism of the density wave instability in 2D dipolar Fermi gases
is arisen from combined effects of the long-range nature of the dipole-dipole interaction
and the Fermi surface effect in the 2D system.
For 3D dipolar Fermi gases, the dipole-dipole interaction does not depend on
the magnitude of momentum~\cite{Miyakawa08} and the Fermi surface effect is less significant.
Thus the density wave phase may not appear in the 3D system.

\begin{figure}
    \begin{center}
    \includegraphics[width=7.cm]{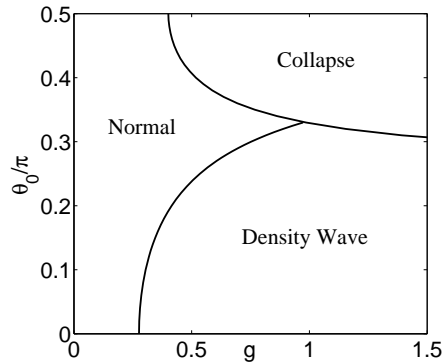}
    \end{center}
    \caption{Phase diagram at zero temperature in terms of $g$ and $\theta_0$.}
    \label{fig7}
\end{figure}

Figure~\ref{fig7} shows a phase diagram at zero temperature in terms of $g$ and $\theta_0$. 
While the density wave instability is identified by Eq.~(\ref{instcond}),
the collapse instability is identified with a negative value of the inverse compressibility, 
$\kappa^{-1}<0$ where
\begin{equation}
   \kappa^{-1} = \frac{2\hbar^2 n_{2d}^2}{m}\left[ \mathcal{E}_{tot}+
    \frac{7}{8} g \frac{\partial \mathcal{E}_{tot}}{\partial g}
    +\frac{1}{8}g^2\frac{\partial^2  \mathcal{E}_{tot}}{\partial g^2}
    \right].
\end{equation}
This result reveals that the density wave phase transition takes place
in a broad region for larger $g$ and smaller $\theta_0$ where the system is stable against collapse.

\section{Critical temperature of the density wave phase transition}

In this section,  we apply our analysis for RPA stability into dipolar Fermi gases at finite temperatures
and obtain the critical temperature of the density wave phase transition.

To do so, we judge the stablity condition~(\ref{RPAstability}) with the density-density response function~(\ref{response})
using the single particle energy calculated by the selfconsistent HF equation~(\ref{constraint})
at a finite $T$ without the variational ansatz.
A critical temperature of the phase transition, $T_c$, is a highest temperature at which the condition
\begin{equation}
\label{critical}
   V_{2d}\left(q, \frac{\pi}{2}\right) \chi\left(q, \frac{\pi}{2} ; T_c\right) = -1
\end{equation}
is fulfilled.

\begin{figure}
    \begin{center}
    \includegraphics[width=9cm]{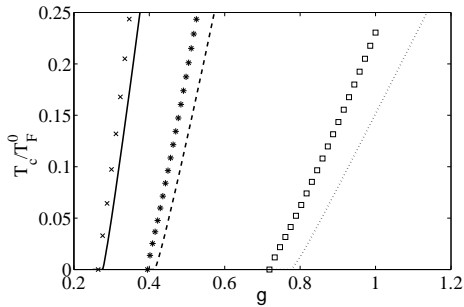}
    \end{center}
    \caption{Critical temperature $T_c$ with the HF single particle energy as a function of $g$ for 
    $\theta_0=0$ (cross), $\theta_0=0.2\pi$ (asterisk),
    and $\theta_0=\arccos(1/\sqrt{3})$ (square).
    The solid line, dashed line, dotted line are $T_c$ with the VM single particle energy
    for $\theta_0=0, 0.2\pi, \arccos(1/\sqrt{3})$, respectively.}
    \label{fig8}
\end{figure}

\begin{figure}
    \begin{center}
    \includegraphics[width=8cm]{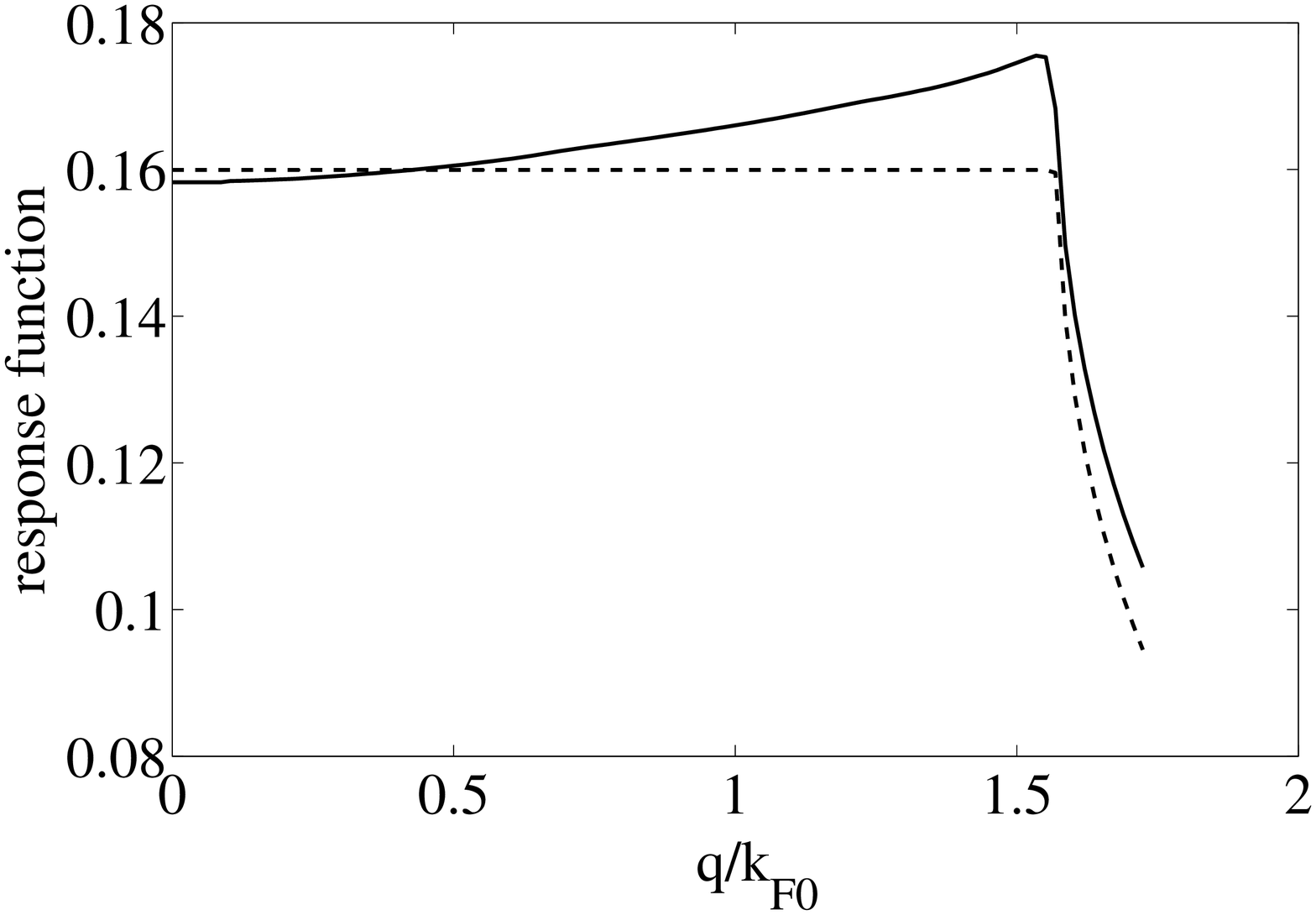}
    \end{center}
    \caption{Density-density response function in units of $m/\hbar^2$ for $g=0.928$, $\theta_0=\arccos(1/\sqrt{3})$,
    and $T=0.02T_F^0$ in HF approximation (solid line) and VM (dashed line).}
    \label{fig9}
\end{figure}

Figure~\ref{fig8} shows $T_c$ in units of ideal gas Fermi temperature 
$T_{F}^0\equiv \hbar^2k_{F0}^2 / 2mk_B$, as a function $g$ for $\theta_0=0$ (cross),
$\theta_0=0.2\pi$ (asterisk), and $\theta_0=\arccos(1/\sqrt{3})$ (square).
In Fig.~\ref{fig8}, we also plot results by replacing $\epsilon(\bm k)$ in Eq.~(\ref{Fermidistribution})
with the VM single particle energy of Eq.~(\ref{VMSPE}) where $\lambda$ and $\alpha_0$
are results of the corresponding system at zero temperature.
The critical temperature with the HF single particle energy is higher than that with
the VM single particle energy for same $g$ and $\theta_0$.
This indicates that the response function $\chi(q,\pi/2)$ at around $q= k_F(\pi/2)$ in HF approximation
is larger than that in VM as shown in Fig.~\ref{fig9}.
This is a consequence of overestimate at $k> k_F(\pi/2)$ for the VM single particle energy, as shown in Fig.~\ref{fig5},
which leads to decrease of the response function via the denominator of integrand in Eq.~(\ref{response}).

It is well known that the charge density wave instability in 1D conductor is due to the nesting property 
that is the increase of the number of lower energy excitations of particle-hole pairs near the Fermi surface~\cite{Gruner}.
The fact that the curvature of the single particle energy at $\phi=\pi/2$ in HF approximation is lower than that in VM
brings a large enhancement of the nesting property for a highly deformed Fermi surface.
As shown in Fig.~\ref{fig8}, the difference between two results is remarkable for larger $\theta_0$. 
Thus we conclude that the deformation of the Fermi surface furthers the density wave phase transition.

\begin{figure}
    \begin{center}
    \includegraphics[width=7cm]{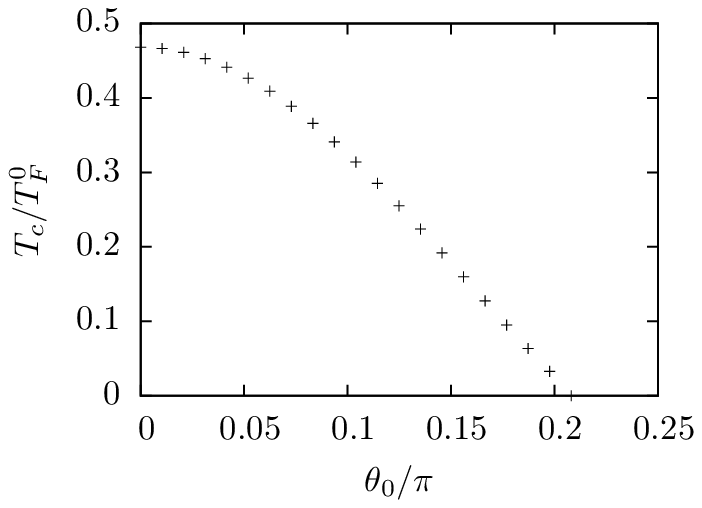}
    \end{center}
    \caption{Critical temperature $T_c$ as a function of $\theta_0$ for $g=0.409$.}
    \label{fig10}
\end{figure}

Finally, let us consider how high the critical temperature will be for a realistic system.
For $^{40}$K-$^{87}$Rb polar molecules with the electric dipole moment of $0.566$ Debye observed
by JILA group~\cite{Ni08}, $m=127$ a.m.u, and the number density of $2.0\times10^7$ ${\rm cm}^{-2}$, we have $g= 0.409$.
Figure~\ref{fig10} reveals that $T_c$ increases with decreasing $\theta_0$
and can be a significant fraction of $T_F$ .
Such a high critical temperature suggests that the density wave of fermionic polar molecules
can be observed in future experiments.

\section{summary}
In the present paper we have studied a realization of the density wave phase transition in a 2D dipolar Fermi gas
where the dipole moments of fermions are aligned by an orientable external field.
Owing to judge the stability of a normal gas, we have investigated RPA stability against density fluctuations of finite momentum.
We showed that the density wave instability takes place in a broad region where the system
is stable against collapse.
We found that the critical temperature $T_c$ of the phase transition can be achieved at
a significant fraction of $T_F$ for a realistic system of polar molecules.

As discussed in Ref.~\cite{Bruun08}, the superfluid phase transition of $p$-wave Cooper pairs is possible
to realize for $\theta_0 \agt \arcsin(2/3)=0.23\pi$ at $T=0$.
The combination of this and our conclusions offers a challenging quest for a coexisting phase of
the superfluid and density wave orders in a 2D dipolar Fermi gas.
This will be discussed in a future publication.

\acknowledgments 
We are very grateful to Dr. A. Suzuki for his continuous encouragement and useful discussions
at Tokyo University of Science where much of the work was done.
TM acknowledges helpful discussions with Dr. H. Pu and his hospitality at Rice University
where part of the work was completed.
TM also thanks Dr. T. Nikuni for useful comments.
TS is supported by the DFG Grant No. RO905/29-1.

\appendix

\section{calculation of variational energy}

In this appendix, we obtain the variational energy at zero temperature
under the assumption of the variational distribution function~(\ref{defdis}).
The total energy in HF approximation is composed of the kinetic energy $E_{kin}$ and interaction energy $E_{int}$
defined by Eqs.~(\ref{ekin}) and (\ref{eint}), respectively.

First, the kinetic energy is easily obtained by
\begin{equation}
\label{A1}
   E_{kin} = \frac{\hbar^2 n_{2d}^2\pi}{2m} \left( \frac{1}{\alpha^2}+\alpha^2 \right).
\end{equation}

Next, we calculate the interaction energy
\begin{eqnarray*}
   E_{int} &=& -\frac{1}{2} \int\frac{d^2k}{(2\pi)^2} \int \frac{d^2k^\prime}{(2\pi)^2}\, 
  f(\bm{k}) f(\bm{k}') V_{2d}(\bm{k} -\bm{k}') \\
  &=& -\frac{4d^2}{\pi^3}k_{F0}^5 I(\alpha; \theta_0) C,
\end{eqnarray*}
where $I(\alpha; \theta_0)$ is defined by Eq.(\ref{deffunc})
and 
\[
   C = \int^\infty_0 dx x^2 \int_0^\infty dy y \int^{\pi/2}_0 d\phi 
   \Theta(1-x^2-y^2-2xy\cos\phi).
\]
We define the function $f(\phi)$ as
\[
   f(\phi) = 1-x^2-y^2 - 2xy\cos\phi\geq f_{min}
\]
where $f_{min}= 1-(x+y)^2$.
If $f_{min} \geq 0$ then $\Theta(f(\phi)) =1$ for $0\leq\phi\leq \pi/2$,
otherwise $\Theta(f(\phi)) =1$ for $\phi_0 \leq \phi \leq \pi/2$ and
$\Theta(f(\phi)) =0$ for $0 \leq \phi < \phi_0$ where
the angle $\phi_0$ is determined by $f(\phi_0)=0$, that is
\[
   \phi_0= \arccos\left(\frac{1-x^2-y^2}{2xy}\right).
\]
With $C=C_1-C_2$ where
\[
   C_1=\frac{\pi}{2} \int^1_0 dx\, x^2 \int^{\sqrt{1-x^2}}_0 dy\, y=\pi/30 ,
\]
and
\begin{eqnarray*}
   C_2&=&\int^1_0 dx \, x^2 \int^{\sqrt{1-x^2}}_{1-x}dy \, y \arccos\left(\frac{1-x^2-y^2}{2xy}\right) \\
   &=& \pi/30-2/45,
\end{eqnarray*}
we obtain the interaction energy as
\begin{equation}
\label{A2}
   E_{int}= -\frac{32}{15}\frac{\hbar^2 n_{2d}^2}{m} g I(\alpha;\theta_0).
\end{equation}
Combining Eq.~(\ref{A1}) and Eq.~(\ref{A2}), we obtain the variational energy as
\begin{equation}
   E_{tot}= \frac{\hbar^2 n_{2d}^2}{m} \left[\frac{\pi}{2}\left( \frac{1}{\alpha^2}+\alpha^2 \right)
   -\frac{32}{15}g I(\alpha;\theta_0) \right].
\end{equation}



\begin{references}
\bibitem{Carr09} L.~D.~Carr, D.~Demille, R.~V.~Krems and J.~Ye, New J. Phys. {\bf 11}, 055049 (2009).
\bibitem{BaranovReview} M.~A.~Baranov, Phys. Rep. {\bf 464}, 71 (2008).
\bibitem{Ni08} K. -K. Ni, S. Ospelkaus, M.~H.~G de Miranda, A.~Pe'er, B.~Neyenhuis, J.~J.~Zirbel,
S.~Kotochigova, P.~S.~Julienne, D.~S.~Jin, J.~Ye, Science {\bf 322}, 231 (2008). 
%
\bibitem{Miyakawa08} T.~Miyakawa, T.~Sogo, and H.~Pu, Phys.~Rev.~A {\bf 77}, 061603(R) (2008).
\bibitem{Sogo09} T.~Sogo, L.~He, T.~Miyakawa, S.~Yi, and H.~Pu, 
New J. Phys. {\bf 11}, 055017 (2009).
\bibitem{Tohyama09} M.~Tohyama, J.~Phys.~Soc.~Jpn., {\bf 78}, 104003 (2009).
\bibitem{Zhang09} J.-N.~Zhang and S.~Yi, Phys.~Rev.~A {\bf 80}, 053614 (2009).
\bibitem{Nishimura09} T.~Nishimura and T.~Maruyama, arXiv:09071757.
\bibitem{Ronen09} S.~Ronen and J.~Bohn, arXiv:0906.3753.
\bibitem{Chan09} C-K.~Chan, C.~Wu, W-C.~Lee, and S. Das~Sarma, Phys.Rev.~A {\bf 81}, 023602 (2010).
\bibitem{Zhang10} J.~-N.~Zhang, S.~Yi, arXiv:1001.0426.
\bibitem{Kestner10} J.~P.~Kestner, S.~Das~Sarma, arXiv:1001.4763.
\bibitem{Endo10} Y.~Endo, T.~Miyakawa, T.~Nikuni, arXiv:1002.0408.
%
\bibitem{BaranovSF} M.~A.~Baranov, M.~S.~Mar'enko, Val~S.~Rychkov, G.~V.~Shlyapnikov,
Phys.~Rev.~A {\bf 66}, 013606 (2002); M.~A.~Baranov, L.~Dobrek, and M.~Lewenstein,
Phys.~Rev.~Lett. {\bf 92}, 250403 (2004).
\bibitem{Zhao09} C.~Zhao, L.~Jiang, X.~Liu, W.~M.~Liu, X.~Zou, and H.~Pu, arXiv:0910.4775.
\bibitem{Bruun08} G.~M.~Bruun and E.~Taylor, Phys.~Rev.~Lett. {\bf 101}, 245301 (2008).
\bibitem{Cooper09} N.~R.~Cooper and G.~V.~Shlyapnikov, Phys.~Rev.~Lett. {\bf 103}, 155302 (2009).
\bibitem{FregosoNJP} B.~M.~Fregoso, K.~Sun, E.~Fradkin, and B.~L.~Lev,
New J. Phys. {\bf 11}, 103003 (2009).
\bibitem{FregosoPRL} B.~M.~Fregoso and E.~Fradkin, Phys.~Rev.~Lett. {\bf 103}, 205301 (2009).
%
\bibitem{Frohlich54} H.~Fr$\rm \ddot{o}$hlich, Proc.~R.~Soc.~London, Ser. A {\bf 223}, 296 (1954).
\bibitem{Peierls} R.~E.~Peierls, {\it Quantum Theory of Solids} (Oxford University Press, New York, 1955).
\bibitem{Gruner} G.~Gr$\rm \ddot{u}$ner, {\it Density Waves in Solids} (Addison-Wesley Longmans, Reading, MA, 1994).
%
\bibitem{Quintanilla09} J.~Quintanilla, S.~T. Carr, and J. J. Betouras, Phys.~Rev.~A {\bf 79}, 031601(R) (2009).
\bibitem{Lin09} C.~Lin, E.~Zhao, and W.~V.~Liu, Phys.~Rev.~B {\bf 81}, 045115 (2010).
%
\bibitem{2dinteraction} For dipolar fermions trapped in a tight harmonic potential $V(z)$,
the wave function along the $z$-axis is the Gaussian function with the width $d_z=\sqrt{\hbar/m\omega_z}$.
In this case, the effective 2D dipole-dipole interaction is obtained by integrating out $z$-degree of freedom.
The resulting interaction has no divergence and shows the same linear momentum dependence as $V_{2d}(\bm q)$
under the condition $k_{F0} d_z \ll 1$ that has to be fulfilled for effectively 2D Fermi systems.
%
\bibitem{Nozieres} P. Nozieres and D. Pines, {\it The Theory of Quantum Liquid}
(PERSEUS BOOKS, Cambridge, 1999).
\bibitem{noteRPA} In RPA, we neglect the exchange scattering in particle-hold excitations, see Ref.~\cite{Nozieres}.
For the stability condition against density fluctuations of finite momentum ${\bm q}=2k_{F0} \hat{{\bm y}}$, 
the direct scattering of particle-hole excitations dominates over the exchange one 
because of the linear momentum dependence of $V_{2d}(\bm q)$.
\end{references}
\end{document}